\documentclass[graybox]{svmult}

\usepackage{helvet}         
\usepackage{courier}        
\usepackage{type1cm}        
\usepackage{makeidx}         
\usepackage{graphicx}        
\usepackage{multicol}        
\usepackage[bottom]{footmisc}

\usepackage{amsmath}
\usepackage{amssymb}
\usepackage{amsfonts}
\usepackage{hyperref}

\newcommand{\p}{\partial}
\newcommand{\M}{\mathcal{M}}

\begin{document}

\title*{Spacetime Stochasticity and Second Order Geometry}
\author{Folkert Kuipers}
\institute{Folkert Kuipers \at Department of Physics and Astronomy, University of Sussex, Brighton, BN1 9QH, United Kingdom, \email{f.kuipers@sussex.ac.uk}}
%
%
\maketitle

\abstract{We discuss the Schwartz-Meyer second order geometry framework and its relevance to theories of quantum gravity that incorporate a notion of spacetime stochasticity or quantum foam. We illustrate the framework in the context of Nelson's stochastic quantization.}

\section{Introduction}
Since the introduction of the path integral formulation in quantum field theory, stochastic analysis has played a pivotal role in the mathematical construction of quantum field theories \cite{Glimm:1987ng,FKac,NelsonPath}.
Closely related to these developments is the theory of stochastic mechanics which proves equivalences between quantum theories and a specific class of stochastic theories \cite{Guerra:StructAspects,Nelson}. In addition, the stochastic quantization framework used in this theory has proved to be a useful computational tool in the study of quantum field theories \cite{Damgaard:1987rr,Guerra:StructAspects,Parisi:1980ys}.
\par

Here, we argue that the success of stochastic analysis in the study of quantum theories is not limited to flat spaces, but can help to elucidate the interplay between quantum theories and gravity, and could in the future provide  handles in the formulation of a theory of quantum gravity.
\par 

The main argument for this statement is that the tools of stochastic analysis that provide a mathematical basis for Euclidean quantum theories can be extended to the context of pseudo-Riemannian manifolds using second order geometry as developed by Schwartz and Meyer \cite{Emery,Meyer,Schwartz}. Such extensions allow to construct and study physical theories on a fluctuating spacetime or quantum foam.

\section{Dynamics on Manifolds}
We illustrate the framework by considering a particle moving on a $n$-dimensional Riemannian manifold $(\M,g)$. In classical physics, its trajectory is described by a map $x(t): \mathcal{T} \rightarrow \M$, where $\mathcal{T}\subseteq \mathbb{R}$. The trajectory is the solution of the geodesic equation
\begin{equation}
	\ddot{x}^j + \Gamma^j_{kl}(x) \, \dot{x}^k \dot{x}^l = 0,
\end{equation}
which can be rewritten in a first order form for $(x,v)(t): \mathcal{T} \rightarrow T\M$. Alternatively, the velocity can be treated as a vector field on the manifold. In this case the governing equations become
\begin{eqnarray}\label{eq:EQMClass}
	v^j(x) \, \nabla_j v^i(x) &=& 0, \nonumber\\
	\dot{x}^i &=& v^i(x).
\end{eqnarray}
\par

We will now introduce a notion of stochasticity - as e.g. induced by a fluctuating spacetime - in this trajectory. We must thus introduce a probability space $(\Omega,\Sigma,\mathbb{P})$ and promote the position $x$ to a random variable $X:(\Omega,\Sigma,\mathbb{P})\rightarrow(\M,\mathcal{B}(\M),\mu)$ with  $\mu=\mathbb{P}\circ X^{-1}$. This allows to study continuous semi-martingale processes $\{X_t:t\in\mathcal{T}\}$, i.e. $X_t=C_t+M_t$ with $C_t$ a c\`adl\`ag process and $M_t$ a local martingale.
\par

In a stochastic theory, one would then like to derive a set of governing stochastic differential equations, similar to the set of ordinary differential equations in the deterministic theory. These stochastic differential equations should then be interpreted either in the sense of It\^o or Stratonovich. However, the formulation of such systems on manifolds is complicated due to the presence of a non-vanishing quadratic variation
\begin{equation}
	[[X^i,X^j]] 
	:= 
	\lim_{k\rightarrow \infty} \sum_{[t_l,t_{l+1}]\in\pi_k} 
	\left( X^i_{t_{l+1}} - X^i_{t_l}\right) 
	\left(X^j_{t_{l+1}} - X^j_{t_l}\right).
\end{equation}
In the It\^o formulation this quadratic variation leads to a violation of the Leibniz rule. Indeed, for functions $f,g:\M\rightarrow\mathbb{R}$, one obtains a modified Leibniz rule of the form
\begin{equation}
	d_2(f \, g) = f \, d_2g + g \, d_2f + 2\, df \cdot dg,
\end{equation}
where
\begin{eqnarray}
	d_2f &=& \p_i f \, dX^i + \frac{1}{2} \p_i\p_j f\, d[[X^i,X^j]],\nonumber\\
	df \cdot dg &=& \frac{1}{2} \, \p_i f \, \p_j g \, d[[X^i,X^j]].
\end{eqnarray}
\section{Second Order Geometry}
The violation of Leibniz' rule implies that many notions from ordinary differential geometry are no longer applicable in a stochastic framework. However, this can be resolved by extending to second order geometry \cite{Emery,Meyer,Schwartz}.
\par

In second order geometry, first order tangent spaces $T\M$ are extended to second order tangent spaces $T_2\M$ such that a second order vector $V$ can be represented in a local coordinate frame as $V = v^\mu \p_\mu + v^{\mu\nu} \p_\mu\p_\nu$. Similarly, one can construct second order forms $\Omega\in T_2^\ast\M$, which in a local coordinate system are given by $\Omega = \omega_\mu d_2x^\mu + \omega_{\mu\nu} dx^\mu \cdot dx^\nu$.
\par

The link between second order geometry and stochastic motion can now be made explicit by constructing second order vectors as
\begin{eqnarray}
	v^\mu(x) &=& \lim_{h\rightarrow 0} \frac{1}{h} \mathbb{E} \left[X^\mu_{\tau+h} - X^\mu_{\tau} \Big| X_\tau=x \right],\\
	v^{\nu\rho}(x) &=& \lim_{h\rightarrow 0} \frac{1}{2h} \mathbb{E} \left[\left( X^\nu_{\tau+h} - X^\nu_\tau \right) \left( X^\rho_{\tau+h} - X^\rho_{\tau} \right) \Big| X_\tau=x \right].\nonumber
\end{eqnarray}
Here, the first order part is constructed as usual, albeit using a conditional expectation, while the second order part reflects the non-vanishing quadratic variation of the stochastic process. It is important to note that when regarded as a second order vector, $v^{^\mu_{\nu\rho}}$ does not transform covariantly. However, one can recover covariance by constructing contravariant vectors $\hat{v}^{^\mu_{\nu\rho}}$ such that
\begin{eqnarray}
	\hat{v}^\mu &:=& v^\mu + \Gamma^\mu_{\nu\rho} v^{\nu\rho},\nonumber\\
	\hat{v}^{\nu\rho} &:=& v^{\nu\rho}.
\end{eqnarray}
In a similar fashion, one can construct covariant forms $\hat{\omega}_{^\mu_{\nu\rho}}$ by
\begin{eqnarray}
	\hat{\omega}_\mu &:=& \omega_\mu,\nonumber\\
	\hat{\omega}_{\nu\rho} &:=& \omega_{\nu\rho} - \Gamma^\mu_{\nu\rho} \omega_\mu.
\end{eqnarray}
\section{Lie Derivatives and Killing Vectors}
It is possible to generalize many notions from first order geometry to the second order geometry framework, cf. e.g. \cite{Emery,Huang:2022,Kuipers:2021jlh}. One way of doing so is by using the fact that a $n$-dimensional manifold equipped with a second order geometry can be mapped bijectively onto a $n$-dimensional brane embedded in a $\frac{n(n+3)}{2}$-dimensional manifold equipped with first order geometry \cite{Kuipers:2021jlh}.
\par

Here, we focus on the construction of a Lie derivative in second order geometry \cite{Huang:2022,Kuipers:2021jlh}. The second order Lie derivative of a scalar is given by
\begin{eqnarray}
	\mathcal{L}_V f = V f &=& v^\mu \p_\mu f + v^{\mu\nu} \p_\mu \p_\nu f. 
\end{eqnarray}
It is also possible to construct a Lie derivative of a second order vector $U$ along a second order vector $V$. However, this requires that the second order parts of the vector fields are scalar multiples of each other. The Lie derivative is then given by the commutator
\begin{equation}
	\mathcal{L}_V U = [V,U].
\end{equation}
Let us now turn to the construction of a Lie derivative of a first order $(k,l)$-tensor along a second order vector field. The result is a first order $(k,l)$-tensor given by
\begin{equation}
	\mathcal{L}_V T = \mathcal{L}_{\mathcal{F}(V)} T + v^{\mu\nu} \left(\nabla_\mu \nabla_\nu + \mathcal{R}^{\cdot}_{\;\mu\cdot\nu} \right) T,
\end{equation}
where the first part denotes an ordinary first order Lie derivative along a first order vector field, as $\mathcal{F}:T_2\M\rightarrow T\M$ s.t. $V \mapsto \hat{v}^\mu \p_\mu$. Moreover,
\begin{equation}
	\mathcal{R}^{\,\cdot\,}_{\;\;\alpha\,\cdot\,\beta} T^{\mu_1...\mu_k}_{\nu_1...\nu_l}
	=
	\sum_{i=1}^k \mathcal{R}^{\mu_i}_{\;\;\alpha\lambda\beta} T^{\mu_1 ... \mu_{i-1} \lambda \mu_{i+1} ... \mu_k}_{\nu_1...\nu_l}
	- \sum_{j=1}^l \mathcal{R}^{\lambda}_{\;\;\alpha\nu_j\beta} T^{\mu_1 ... \mu_k}_{\nu_1 ... \nu_{j-1} \lambda \nu_{j+1} ... \nu_l}.
\end{equation}
\par 

The construction of Lie derivatives of tensors along second order vector fields allows to construct a notion of a second order Killing vector. We find
\begin{equation}
	\mathcal{L}_K g_{\mu\nu} 
	= \nabla_\mu \hat{k}_\nu + \nabla_\nu \hat{k}_\mu - 2 \, \hat{k}^{\rho\sigma} \mathcal{R}_{\mu\rho\nu\sigma},
\end{equation}
setting this to $0$ leads to the second order Killing equation
\begin{equation}
	\nabla_{(\mu} \hat{k}_{\nu)} = \hat{k}^{\rho\sigma} \mathcal{R}_{\mu\rho\nu\sigma},
\end{equation}
We thus find that a first order killing vector $k^\mu$ must be promoted to the covariant first order part of a second order vector $\hat{k}^\mu$. Secondly, a second order Killing vector has a non-vanishing divergence proportional to the curvature of space. A classical observer will interpret this deviation as a symmetry breaking of the classical spacetime due to the fluctuations.
\section{Stochastic Dynamics on Manifolds}
After setting up the machinery of second order geometry, one can derive stochastic differential equations of motion on a manifold. We will consider a Brownian motion, which is uniquely characterized by the quadratic variation
\begin{equation}
	d[[X^i,X^j]]_t = \alpha \, g^{ij} (X_t) \, dt 
\end{equation}
with $\alpha\in[0,\infty)$. The system given in Eq.~\eqref{eq:EQMClass} now becomes \cite{Kuipers:2021jlh,Kuipers:2021ylr,Nelson}
\begin{eqnarray}
	\left[ g_{ij} \hat{v}^k \nabla_k + \frac{\alpha}{2} \Big( g_{ij} \Box - \mathcal{R}_{ij}\Big) \right] \hat{v}^j 
	&=& \frac{\alpha^2}{12} \nabla_i \mathcal{R}, \nonumber\\
	dX^i &=& v^i \, dt + dM^i,
\end{eqnarray}
which should be interpreted as a system of stochastic differential equations in the sense of It\^o.
\par

As we have only discussed non-relativistic processes on Riemannian manifolds, while the physical world is relativistic, we must extend our discussion to relativistic processes on Lorentzian manifolds. Extensions of second order geometry to Lorentzian manifolds are straightforward, as the framework is developed for any smooth manifold with a connection \cite{Emery}. Furthermore, similar to a classical relativistic theory, the formulation of a relativistic theory on Lorentzian manifolds introduces a relativistic constraint equation \cite{Kuipers:2021aok,Kuipers:2021ylr}. The velocity field is then a solution of the system
\begin{eqnarray}
	\left[ g_{\mu\nu} \hat{v}^\rho \nabla_\rho + \frac{\alpha}{2} \Big( g_{\mu\nu} \Box - \mathcal{R}_{\mu\nu}\Big) \right] \hat{v}^\nu 
	&=& \frac{\alpha^2}{12} \nabla_\mu \mathcal{R}, \nonumber\\
	g_{\mu\nu} \hat{v}^\mu \hat{v}^\nu + \alpha \nabla_\mu \hat{v}^\mu - \frac{\alpha^2}{6}\mathcal{R} 
	&=&  \epsilon
\end{eqnarray}
with $\epsilon\in\{-1,0,+1\}$ for respectively time-like, light-like and space-like particles.
Moreover, after splitting the tangent bundle in time-like, light-like and space-like segments, one can construct a positive definite non-degenerate metric $g_{\rm Eucl.}$, on these segments using a Wick rotation. The stochastic motion is then given by the solution of the It\^o system \cite{Dohrn:1985iu,Kuipers:2021aok}
\begin{eqnarray}
	dX^\mu &=& v^\mu \, d\tau + dM^\mu,\nonumber\\
	d[[X^\mu,X^\nu]] &=& \alpha \, g_{\rm Eucl.}^{\mu\nu} \, d\tau.
\end{eqnarray}
\par

One might object that we have only discussed a classical Brownian motion which is not obviously related to quantum mechanics. However, one can complexify the manifold and study processes satisfying
\begin{equation}
	[[Z^\mu,Z^\nu]]_\tau = \alpha \, g^{\mu\nu} (Z_\tau) 
\end{equation}
with $Z=X+{\rm i} \, Y$ and $\alpha\in\mathbb{C}$. Then, for the choice $\alpha=\frac{\rm{i} \, \hbar}{m}$, the real projection of this process describes a free scalar quantum particle with mass $m$ on the manifold \cite{Kuipers:2021ylr}.
\par

\section{Conclusions \& Outlook}
We have discussed the second order geometry framework and shown that it allows to describe stochastic dynamics on manifolds. Moreover, we have discussed extensions to relativistic stochastic dynamics and discussed the close relation between stochastic and quantum dynamics.
\par

We should note that in this paper we have only described free particles moving in a fixed geometry. Although this picture can be extended to include external forces derived from scalar or vector potentials and the notion of spin, see e.g. \cite{Nelson}, a field theoretic formulation will be necessary to consider dynamical geometries, and to study quantum aspects of gravity. Stochastic field theories have been discussed in the context of Nelson's stochastic quantization, cf. e.g. Ref.~\cite{Guerra:StructAspects}, but the subject is not yet as mature as it is in the Parisi-Wu formulation of stochastic quantization \cite{Damgaard:1987rr,Parisi:1980ys}.
\par

Nevertheless, the discussion of point particles presented in this paper provides an indication of the geometrical structure that is necessary to formulate such theories. Indeed, the configuration space of a classical particle is the tangent bundle $T\M$, which, in the stochastic framework, is promoted to a second order tangent bundle $T_2\M$. The configuration space of a classical field theory, on the other hand, is a first order jet bundle $J^1\pi$ over the manifold $\M$. It is thus expected that the configuration space for a stochastic field theory is a second order jet bundle $J^2\pi$. We note that the construction of classical field theories on higher order jet bundles has been discussed in the literature~\cite{Campos:2009ue,Campos:2010ay}.

\begin{acknowledgement}
This work is supported by a doctoral studentship of the Science and Technology Facilities Council, award number: 2131791.
\end{acknowledgement}
\end{document}